# Comment on "Investigations into the impact of astronomical phenomena on the terrestrial biosphere and climate" (arXiv:1505.07856 [astro-ph.EP]) by Fabo Feng.

Adrian L. Melott

This work by Feng and papers which published its conclusions do not cite nor do they deal with objections by the author published in 2013-2014. There are many fundamental problems. We will summarize here the principal problems, as based on Melott and Bambach (2013, 2014) which render irrelevant most of the work presented by Feng. Feng did not cite or deal with either of these papers, nor include the missing, relevant work Melott and Bambach (2010). We restrict our attention to the question of the detection of periodicity. Readers are referred to Melott and Bambach (2013, 2014) for fuller detailed discussion of the problems.

1. Biodiversity is a product of both origination and extinction. Biodiversity increases when origination exceeds extinction and decreases when extinction exceeds origination. The behavior of just one of these phenomena does not, alone determine biodiversity. Feng only examines extinction rates and events and apparently did not deal with biodiversity (even though they used the term in implying their study showed no strong 62 Myr periodicity), whereas this periodicity in biodiversity has been shown to result from the coherent interaction of extinction and origination fluctuations, and neither origination or extinction alone produce a strong signal.

2. Feng ignores the substantial effect wherein the impact of the 62 Myr biodiversity periodicity is found to be diluted in the last 150 Myr by the accumulation of long-lived genera, apparently resistant to the cycles. When long-lived genera are removed, the intensity remains strong up to the Recent. This is not a severe impediment to the detection of extinction periodicity, but it would be if they were actually examining biodiversity, as claimed.

3. There is no evidence that he has used the 2012 geological timescale; our previous work shows that this increases the signal for periodicity over previous standards.

4. Feng asserts that random data may produce a strong periodicity, so finding some periodicity that fits is not significant. However, he does not ask how likely this is given the data and given the strength of the spectral peak, nor do they mention the tests by Rohde & Muller, Cornette, Lieberman, Melott, Bambach, and others.

Furthermore, they assert that they are testing the viability of periodic models by averaging over all such models. This of course dilutes any benefit of hitting the "correct" model—which is actually rather well-specified given past work. In exploratory data analysis (Tukey 1977) the next step from data is the analysis of the data to suggest a model—which is then subjected to hypothesis testing. The guiding principle is the statement by Tukey (1962): "Far better an approximate answer to the right question, which is often vague, than an exact answer to the wrong question, which can always be made precise."

5. Feng has assigned both mass extinction events and "continuous" extinction rates to the middle of the interval in question. However, as noted in Foote (2005) there is considerable evidence that most extinctions are pulsed and concentrated at the end of intervals and it is certainly the case for larger extinction events, because these events are used to set interval boundaries. Therefore it is a better approximation to assign extinctions to the end of intervals than to the midpoints.

6. Feng assigns an uncertainty to the extinction data derived from the length of the interval to which they are assigned. This is almost certainly an overestimate of the true uncertainty in date assignments, since the uncertainty in geological date assignments varies with their date, but is now typically less than ± 1 Myr. Consequently, nearly all the Feng assignment of extinctions are outside the 95% confidence interval for the timing, and more than 2/3 exceed the estimated systematic error." This will result lowered significance in their Bayesian procedure due to large uncertainty in the data.

7. Feng also ignore the finding of periodicities coincident with biodiversity fluctuations within the errors in period and phase in the emplacement of sediments (Meyers and Peters 2011) or the ratio of $^{87}Sr/^{86}Sr$ (Melott et al. 2012). Of course, as a formally constituted statistical question, these are not relevant: Feng were claiming to test for biodiversity signals. But the practical scientific issue is that these variables are known to be related to biodiversity.

8. The method of Akaike weights is a method of converting data on likelihood into a conditional probability for each of a set of possible models into a probability that additional data would support that model. Using numbers previously published by Feng, we found that random draws from a periodic function) has a probability of 90% of being the "best" model of the 19 they considered. Clearly their analysis, using the maximum likelihood method evaluated as a probability, shows a preference for a periodic parent distribution behind mass extinction events.